\documentclass[useAMS,usenatbib,usegraphicx]{mn2e}
\usepackage{amsmath}
\usepackage[toc,page]{appendix}
\renewcommand{\vec}[1]{{\bf{#1}}}
\newcommand{\fr}[2]{{\displaystyle \frac{#1}{#2}}}
\newcommand{\diff}[2]{{\fr{d{#1}}{d{#2}}}}
\newcommand{\pdiff}[2]{{\fr{\partial{#1}}{\partial{#2}}}}

\usepackage{pslatex}
\usepackage{amsmath}

\title[Comparative Analysis of the Outflow Model]{Comparative Analysis of the Model for Exoplanet Atmosphere Outflow}
\author[Isakova et. al.]{P. B. Isakova,
        Ya. N. Pavlyuchenkov,
        E. S. Kalinicheva, and
        V. I. Shematovich \\
        Institute of Astronomy RAS, Moscow, Russia}
\date{}
\begin{document}
\date{Received November 17, 2020; revised January 22, 2021; accepted January 29, 2021 \\
isakovapb@inasan.ru}
\pagerange{445--454} \pubyear{2021}
\volume{65}

\maketitle

\begin{abstract}
\noindent
Modeling the outflow of planetary atmospheres is important for understanding the evolution of exoplanet systems and for interpreting their observations. Modern theoretical models of exoplanet atmospheres become increasingly detailed and multicomponent, and this makes difficulties for engaging new researchers in the scope. Here, for the first time, we present the results of testing the gas-dynamic method incorporated in our aeronomic model, which has been proposed earlier. Undertaken tests support the correctness of the method and validate its applicability. For modeling the planetary wind, we propose a new hydrodynamic model equipped with a phenomenological function of heating by stellar UV radiation. The general flow in this model well agrees with results obtained in more detailed aeronomic models. The proposed model can be used for both methodical purposes and testing the gas-dynamic modules of self-consistent chemical-dynamic models of the planetary wind.\\\\
{\bf DOI:} 10.1134/S1063772921060032
\end{abstract}

\def\gtrsim{\mathrel{\hbox{\rlap{\hbox{\lower5pt\hbox{$\sim$}}}\hbox{$>$}}}}

\section{Introduction}
The discovery of exoplanets undoubtedly is among the most significant achievements of modern astronomy. Over the past decades, we have progressed a long way from hypotheses on the natural occurrence of exoplanets to revealing several thousand extrasolar planets (\cite{2017JGRE..122...53D}, https://exoplanetarchive.ipac.caltech.edu/). The Kepler Cosmic Telescope has found unexpected configurations of planetary systems, which differ from the Solar System. The distribution of exoplanet radii determined at transit observations is continuous~\cite{2015ApJS..217...16R}, and varying from planets of the terrestrial type with sizes $<1~R_\otimes$ to gigantic planets with radii $>4~R_{\rm J}$. Numerous exoplanets can have rock cores surrounded by thin envelopes of hydrogen and helium
gases. Recent analyses of low-density exoplanets similar to Neptune show that these planets cannot retain their hydrogen envelopes due to extremely high gas-
dynamic rates of atmospheric losses, if the current estimates of their large radii or small masses \cite{2017MNRAS.466.1868C, 2017A&A...598A..90F}. are accurate enough. For these findings, we can argue that the estimates used sometimes underestimate masses or overestimate planet radii (due to clouds at high altitudes). Alternatively, we can conclude that hot Neptunes demonstrate higher albedos than those of exoplanets similar to Jupiter \cite{2017MNRAS.466.1868C}.

Exoplanetary atmospheres are very multifarious. These atmospheres are key objects for understanding the physics and general features of exoplanets. For example, it was unexpected to discover several rocky planets of low masses with envelopes of light gases of H$_2$ and/or H$_2$O. This indicates that numerous planets, which are considered as those of the terrestrial type, can partially retain their initial protoatmospheres of hydrogen and helium \cite{2014MNRAS.439.3225L,2015NatGe...8..177T,2015AsBio..15...57L, 2016SSRv..205..153M}. It advances the modeling of observable features of exoplanet atmospheres as a whole and their biomarkers in particular, since life on exoplanets is a global topic of research.

Researches on planetary exospheres cannot be successful without studying thermospheres and ionospheres, which are usually called the upper atmospheres. The structure and features of upper atmospheres clarify the effects of parental stars. Under strong stellar UV radiation, the upper planetary atmosphere can be expanded over large distances from the center of the planet. This results in rapid atmospheric losses, with their efficiency being determined by the physics and chemistry of the atmosphere \cite{2008SSRv..139..355J,2015ASSL..411..105S, 2018PhyU...61..217S}. For example, the so-called gas-dynamic outflow (or the planetary wind) was studied theoretically for planets of the Solar System at early stages of their evolution \cite{1981Icar...48..150W,1987Icar...69..532H, 1996JGR...10126039C, 2016MNRAS.459.2030V}.Only recent decades have proposed opportunities for observing the gas-dynamic outflow from several close exoplanets \cite{ 2003Natur.422..143V, 2004ApJ...604L..69V, 2010ApJ...717.1291L, 2014ApJ...786..132K}. Some attempts on numerical modeling of these phenomena were undertaken \cite{2004Icar..170..167Y,
2007P&SS...55.1426G,2009ApJ...693...23M,2013Icar..226.1678K, 2014ApJ...795..132S, 2017ARep...61..387I, 2019AREPS..47...67O}. In particular, three-dimensional effects of the stellar wind and the magnetic field on the structure and observable features of exoplanets have been studied \cite{2018MNRAS.475..605C, 2019ARep...63..550Z}. 

All these works indicate that theoretical studies of planetary atmospheres are successfully progressing. At the same time, it is difficult to explain the fundamentals of the theory of atmospheres for entry-level researchers. The problem of dynamics of spherically symmetric atmospheres, which are closely related to
the dissipation of gaseous envelopes, presents a classical example of the sort. This problem is similar to that of the stellar wind solved analytically in the early 1960s~\cite{1960ApJ...132..821P, 1964ApJ...139...72P}. However, the planetary atmospheres demonstrate some specific features related to their heating and cooling processes. Studying the formation and evolution of primary and secondary atmospheres and the potential habitability of exoplanets is of paramount significance for a number of modern scientific problems including those on the cosmogony of the Solar System and the origin of life on Earth. This is especially important for planets of the terrestrial type (sub-, exo-, and super-Earths) and for planets of the sub-Neptune type, as well as for ocean-planets which have no counterparts in the Solar System.

The purpose of the present work is to test a method used earlier for calculations on the outflow of exoplanet atmospheres~\cite{2017ARep...61..387I,2018MNRAS.tmp..609I}. Here, we analyze the gas-dynamic modeling of isothermal atmospheres and present a model of the atmosphere outflow with a simple heating function used to reproduce the atmospheric losses on exoplanets. Our work will be helpful for testing gas-dynamic methods applied to modeling the planetary wind.

\section{Testing the Gas-Dynamic Method}

Modeling the dynamics of exoplanet atmospheres is based on solutions of the set of gas-dynamic equations
\begin{equation}\label{eq-density}
 \pdiff{\rho}{t} + \nabla \cdot \left( \rho {\vec v} \right) = 0,
\end{equation}
\begin{equation}\label{eq-velocity}
 \pdiff{{\vec v}}{t} + \left( {\vec v} \cdot \nabla \right) {\vec v} =
 -\frac{\nabla P}{\rho} - \fr{G M {\vec r}}{r^3},
\end{equation}
\begin{equation}\label{eq-energy}
 \pdiff{\varepsilon}{t} +
 \left( {\vec v} \cdot \nabla \right) \varepsilon =
 - \fr{P}{\rho} \nabla \cdot {\vec v} + \Gamma,
\end{equation}
where $\rho$~--- the density, ${\vec v}$~--- velocity, $P$~--- pressure, $G$~--- gravitation constant, $M$~--- planet mass, $\varepsilon$~--- specific thermal
energy (per unit mass), $\Gamma$~---heating-cooling function calculated per unit mass. Here, we assume that the atmosphere mass is small compared with the planet mass, i.e., the self-gravitation of the atmosphere is neglected. The density, temperature, and pressure are interrelated by the ideal gas law 
\begin{equation}\label{eq-state}
 P = nkT,
\end{equation}
where $n = \rho / m$ is the number density of molecules, $m$~--- molecular mass, $k$~--- Boltzmann constant, $T$~--- temperature, while the thermal energy and the temperature are related by the equation
\begin{equation}\label{eq-state-energy}
 \varepsilon = \fr{i}{2}\fr{k T}{m} = \fr{1}{\gamma-1}\fr{k T}{m} ,
\end{equation}
where $i$~--- the number of degrees of freedom for molecules, and $\gamma$ --- adiabatic index. Further, we assume a spherical symmetry of the atmosphere, which permits analytical solutions of the problem.

For nonzero heating-cooling functions, the stationary atmospheric outflow can be described by the following equations for the variables $v$ and $\varepsilon$:
\begin{eqnarray}
 && \left[ v - \left( \gamma - 1 \right) \fr{\varepsilon}{v} \right] \diff{v}{r} + \left( \gamma - 1 \right) \diff{\varepsilon}{r} = \left( \gamma - 1 \right) \fr{2 \varepsilon}{r} - \fr{G M}{r^2}\\
 && v \diff{\varepsilon}{r} + \left( \gamma - 1 \right) \varepsilon \diff{v}{r} = - \left( \gamma - 1 \right) \fr{2 \varepsilon v}{r} + \Gamma.
\end{eqnarray}
For realistic heating-cooling functions, obtaining analytical solutions of these equations becomes rather difficult. A numerical integration of this set of equations
(for example, using the Runge--Kutta method) is also difficult due to possible critical points (beforehand unknown). Another problem of the numerical integration is in choosing the internal boundary condition corresponding to physically stable solutions. By simple transformations, we can find an alternative set of equations
\begin{eqnarray}
&&\diff{}{r} \left[ \fr{v^2}{2} + \gamma \varepsilon - \fr{G M}{r} \right] = \fr{\Gamma}{v} \\
&&\diff{}{r} \left[ \varepsilon v^{\gamma - 1} r^{2 \gamma - 2} \right] = \fr{\Gamma r^{2 \gamma - 2}}{v^{2 - \gamma}}.
\end{eqnarray}
A specific feature of these equations for ? = 0 , i.e., for adiabatic processes, is that the expressions in square brackets are conserved. A difficulty encountered at the numerical integration of these equations is that the nonlinear relations between the expressions in square brackets must be simultaneously solved that implies
finding the roots of corresponding nonlinear equations. In addition, the boundary condition must be chosen in a form providing stable solutions. Thus, for
studying the atmosphere outflow, one frequently uses gas-dynamic modeling that is a direct solution for non-stationary equations~\eqref{eq-density}--\eqref{eq-energy}.

Choosing a proper gas-dynamic method, in turn, is a critical problem, since the method must correctly reproduce characteristic features of solutions and avoid artifacts. For modeling the dynamics of spherically symmetric atmospheres~\cite{2017ARep...61..387I}, a completely conservative implicit Lagrangian method described in~\cite{Samarskii:1992} has been used. This method was successful in solving the problem of the collapse of protostellar clouds~\cite{2015ARep...59..133P}. For the cloud collapse, the method successfully reproduces all specific features of the analytical solution~\cite{diser-Yaroslav}. However, the correctness of this
method's application to problems of the atmosphere outflow has not been proved in~\cite{2017ARep...61..387I}. In the presentsection, we show that the given numerical method successfully solves the problem of the isothermal atmospheric outflow whose exact solution is well known.

The numerical method used employs the finite-difference approximation of the original gas-dynamic equations. This method is Lagrangian, i.e., the gas does not overpass cell boundaries, but the cells themselves move being compressed (or expanded) together with the material. All quantities, except velocities, are specified at the cell center, but velocities are specified at cell boundaries. For our calculations, we use 2001 cells, and the initial grid takes a uniform discretization
on $r$ . The left boundary of the integration region is rigorously fixed, i.e., the velocity at the left boundary is zero. The right boundary of the integration region can move, with the outer pressure (in an additional cell right-adjacent to the boundary) being the boundary condition. The position of the atmosphere boundary
is automatically traced by the coordinate of the last cell. A detailed description of the method can be found in~\cite{2017ARep...61..387I}.

Let us consider an atmosphere of a planet with the mass $M = 0.07 M_{\rm J}$ and the radius $a = 0.35 R_{\rm J}$.These values are close to those of the warm Neptune GJ 436b (see Table~\ref{neptune_par}). Let us take the temperature $T_{0} = 8575~\text{K}$, which corresponds to the dimensionless Jeans parameter $\lambda = 10$, (that is the ratio of the gravitational energy at the inner boundary to the thermal energy neglecting the factor $i/2$)
\begin{equation}\label{eq-lambda}
 \lambda = \left. {\dfrac{G M m}{a}} \middle/ {k T_0}\right. ,
\end{equation}
where $T_0$~--- the temperature of the isothermal atmosphere. Let the density at the inner atmospheric boundary be $\rho_{0} = 10^{-12}~\text{g}/\text{cm}^3$. Introduce a nondimensional density of the form $\eta = \rho / \rho_0$. The left boundary of the integration region is stringently connected to the point $x=1$, where $x = r / a$. At the initial time, the right atmospheric boundary is at the nondimensional distance $x=2$. while the density distribution inside the region $1\le x \le2$ is hydrostatic, except the condition at the boundary, where the pressure in the last cell can differ from that in the surrounding medium.

\begin{figure*}
\hfill
 \includegraphics[width=1\columnwidth]{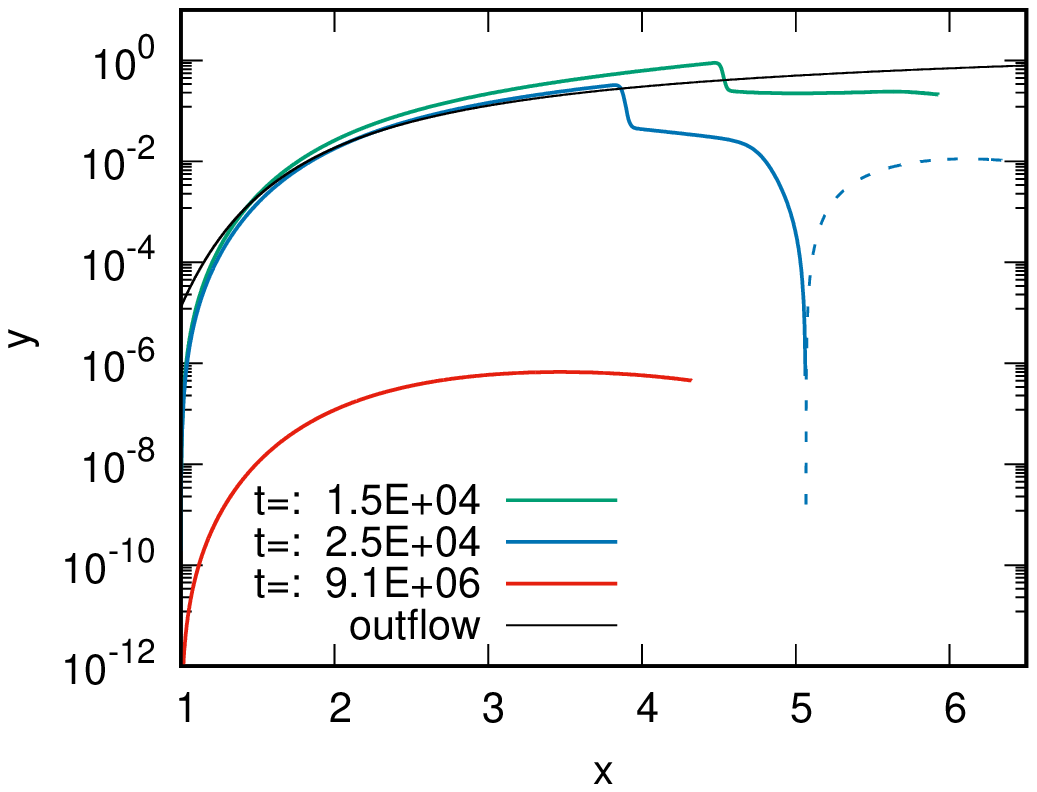}
 \includegraphics[width=1\columnwidth]{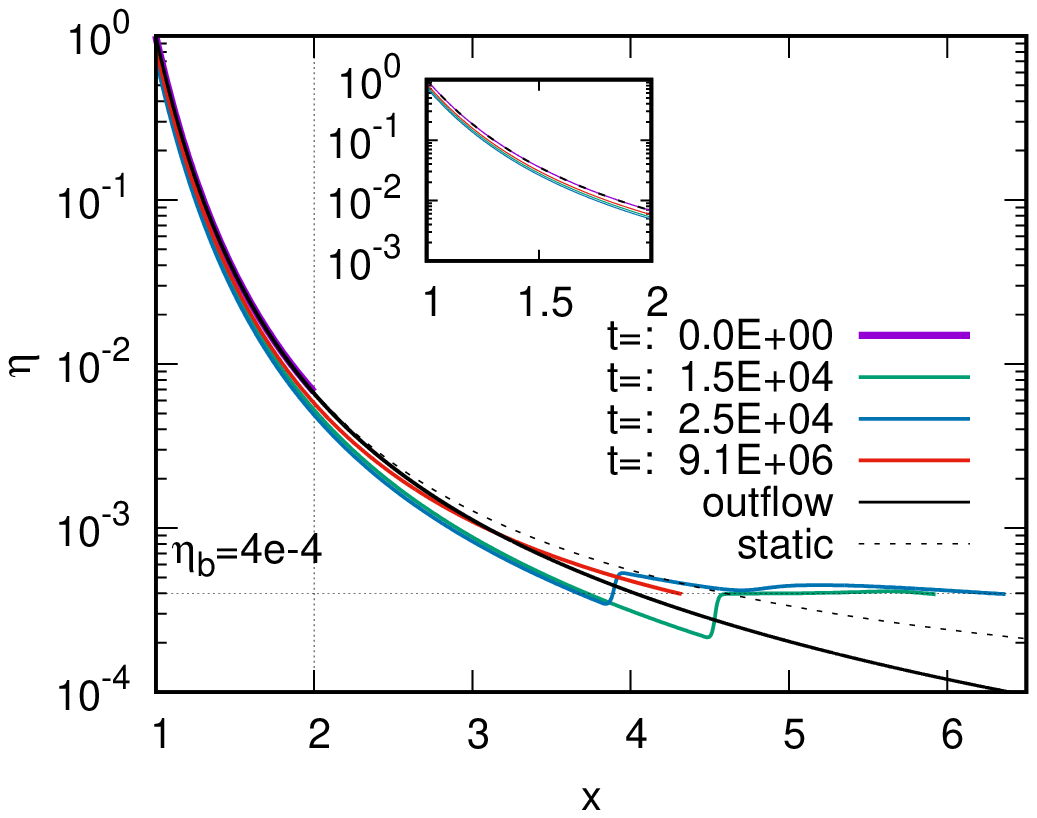} \\
 \hfill
 \includegraphics[width=1\columnwidth]{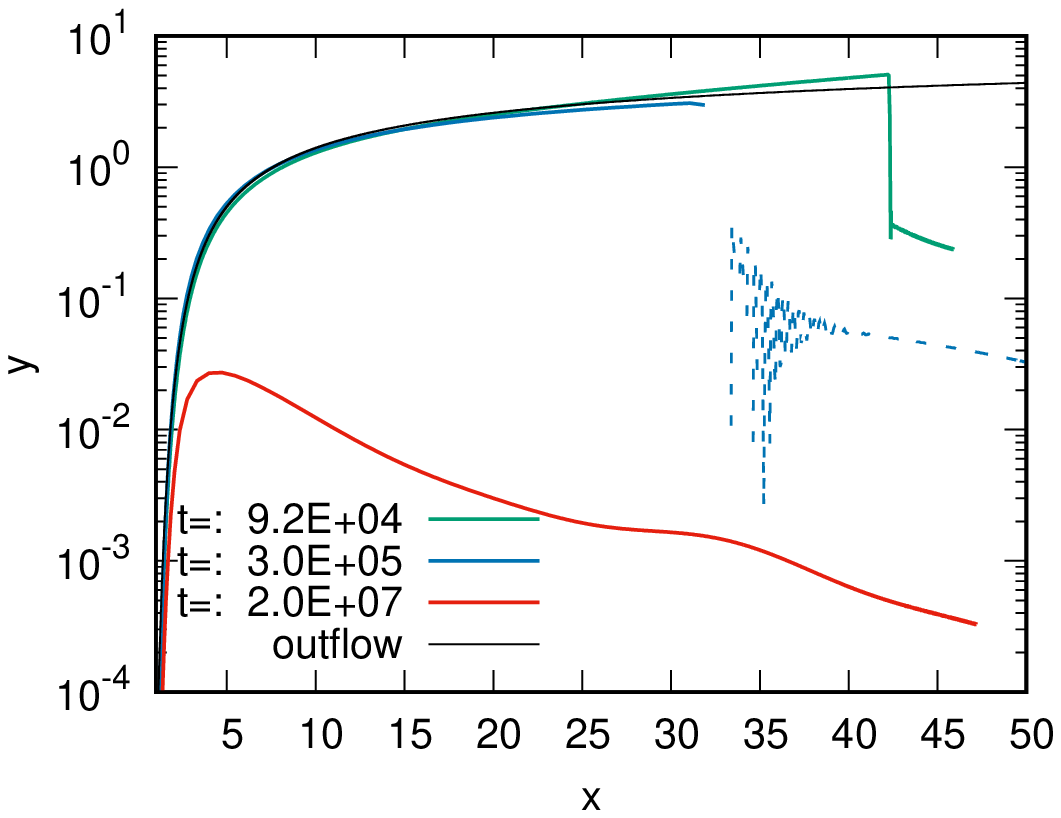}
 \includegraphics[width=1\columnwidth]{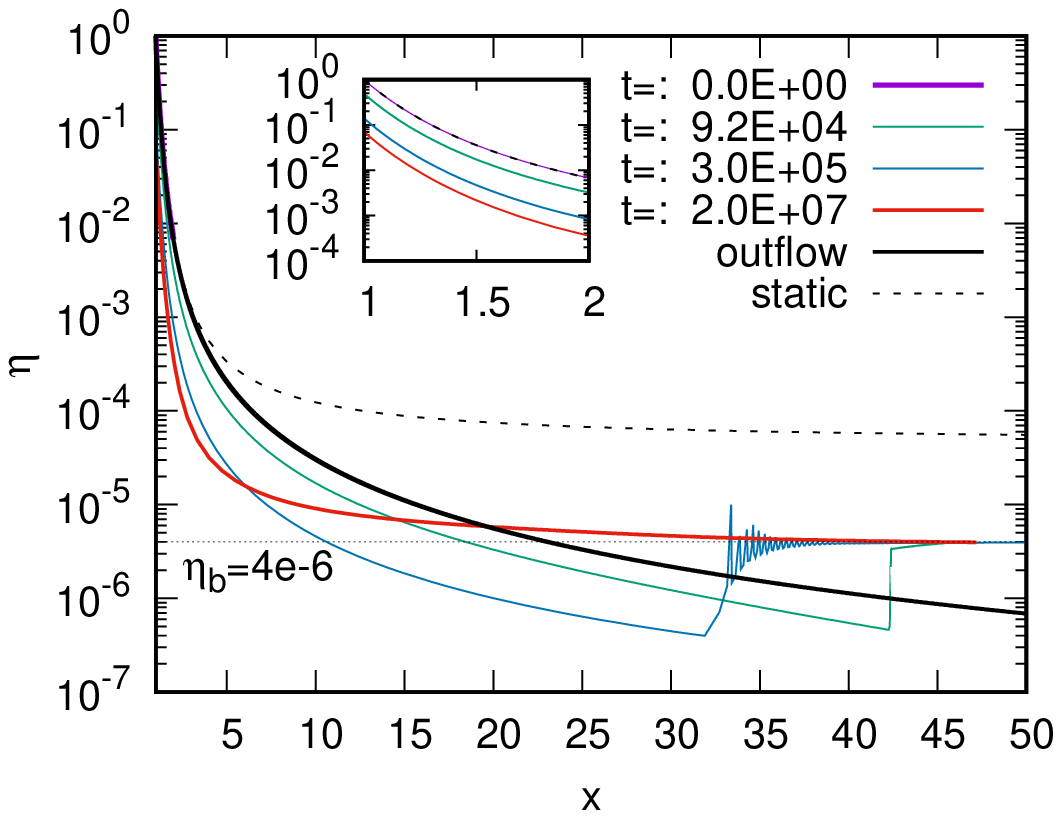} \\
 \caption{Gas-dynamic modeling of isothermal atmospheres with fixed outer boundary conditions for densities. Upper diagrams correspond to the boundary nondimensional density 
$\eta_b = 4\times10^{-4}$, lower diagrams $\eta_b = 4\times10^{-6}$. Dashed curves shown in left panels correspond to negative velocities, solid to positive.}
 \label{isoterm1}
\end{figure*}

Let us consider the calculations executed for several boundary conditions. For the first boundary condition at the right boundary, we take the pressure corresponding to the nondimensional density $\eta_{\rm b} = 4\times 10^{-4}$. This value is lower than the density at the outer boundary of the initial atmosphere ($\eta \approx 7 \times 10^{-3}$ at $x=2$); therefore, the initial distribution must evolve due to the expansion of outer atmospheric regions exposed to the pressure gradient formed at the atmospheric boundary. The value $\eta_{\rm b}$ is by an order of magnitude higher than $\eta_\infty\approx 4.5\times 10^{-5}$, which corresponds to the equilibrium density of the isothermal atmosphere at infinity, $\eta_{\infty}=e^{-\lambda}$. The upper panels of Fig.~\ref{isoterm1} present the dynamics of the given atmosphere numerically calculated in dimensionless variables $\eta$ and $y$, where $y$~--- is the ratio of the kinetic energy of the gas molecule to its thermal energy (neglecting the factor $i/2$):
\begin{equation}\label{eq-y}
 y = \left. \frac{m v^2}{2} \middle/ {k T_0}\right. .
\end{equation} 
At the time of $1.5\times 10^4$~s, the atmosphere expands, with its right boundary reaching $x=6$ and $y$ approaching stationary analytical values. Reaching
$x=6.5$ at the time of $2.5\times 10^4$~s, the atmosphere stops its expansion, and its outermost layers start to compress under an effect of the external pressure. The jumps in distributions $y(x)$ and $\eta(x)$ result from non-equilibrium initial distributions. In what follows, the atmosphere oscillates, with its outer boundary oscillating near $x=4.3$. At later times, the atmosphere gradually reaches its equilibrium, with the final density distribution being very close to the analytical hydrostatic distribution. Thus, for the density boundary conditions $\eta_{\rm b} > \eta_\infty$, the atmosphere reaches its hydrostatic equilibrium, with its boundary corresponding to that $x$,at which the density of the analytical hydrostatic atmosphere is equal to the boundary condition.

If we take the density at the outer boundary below the value $\eta_\infty$, that is $\eta_{\rm b} < \eta_\infty$, the atmosphere will expand to larger radii, but the atmosphere will also reach its equilibrium in the course of time. This is explained by the fact that the Lagrangian method used takes a finite atmospheric mass, while the inner boundary of the integration region is fixed. At the atmosphere expansion, the rarefaction reaches the first cell, and the density in this cell decreases. In the course of time, the density throughout the atmosphere decreases so strongly that the outer boundary condition corresponds to the hydrostatic solution. Fig.~\ref{isoterm1} shows this case in the lower panels, which present a model with the boundary condition $\eta_{\rm b} = 4 \times 10^{-6}$. Note that for intermediate times (see the curves for the time of $9.2\times 10^4$~s in the lower panels of Fig.~\ref{isoterm1}), the profiles $y(x)$ and $\eta(x)$ are close to analytical distributions for the stationary isothermal wind. Oscillations found in the distributions $y(x)$ and $\eta(x)$ can be explained by numerical instabilities inefficiently suppressed with the use of artificial viscosity.

\begin{figure*}
\centering
 \includegraphics[width=1\columnwidth]{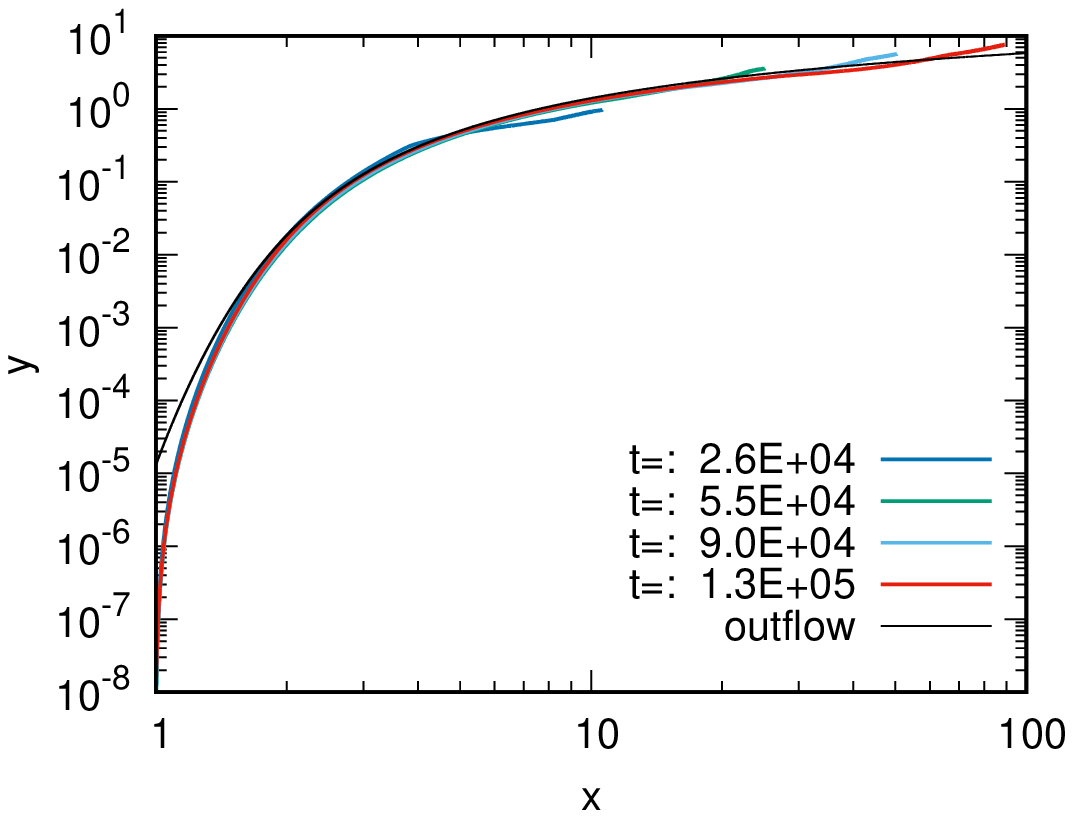} \hfill
 \includegraphics[width=1\columnwidth]{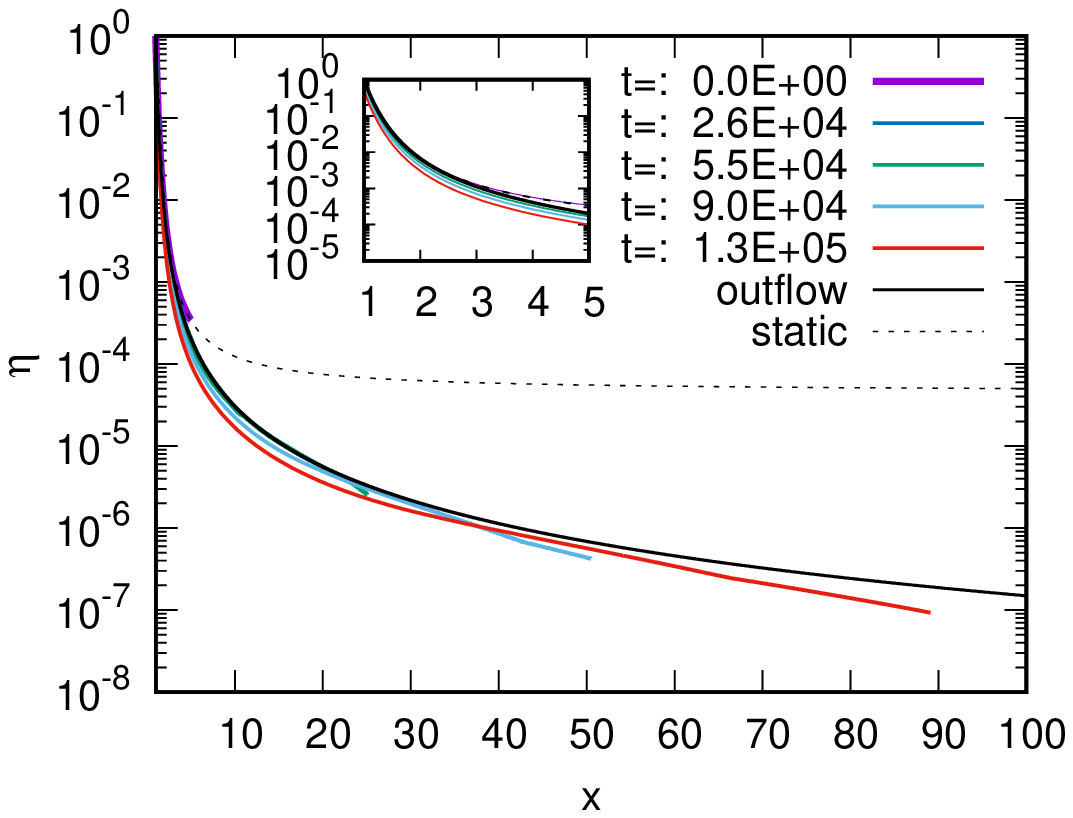} \\
 \caption{Gas-dynamic solution for the case when the density for the atmospheric outer boundary at every time step adaptively changes in accordance with the analytical solution for the stationary outflow.}
 \label{isoterm2}
 \end{figure*}

Fig.~\ref{isoterm2} shows a numerical solution for the case when the density for the atmospheric outer boundary at every time step adaptively changes in accordance
with the analytical solution for the stationary outflow. In other words, $\eta_{\rm b}(x)$ is taken from the analytical solution $\eta(x)$ found for the isothermal wind. In this case, the obtained distributions $\eta(x)$ and $y(x)$ are close toanalytical found for the isothermal wind. The greatest difference between the numerical and analytical solutions is seen in the distribution $y(x)$ near the left atmospheric boundary. This is explained by that the inner atmospheric boundary is fixed, i.e., the velocity at this boundary equals zero. On the other hand, the analytical stationary solution shows for the left boundary nonzero velocities, which correspond to particular $y_0$.

The tests discussed above allow us to argue that the gas-dynamic method is applicable to modeling the atmosphere outflow of exoplanets. Using this method in the next section, we model the outflow of atmosphere with simple heating function.

\section{Atmosphere with Phenomenological Heating Function}

The photoionization and photodissociation of exoplanetary gases by the UV radiation of central stars are the main processes of heating exoplanet atmospheres (see, for example,~\cite{2017ARep...61..387I}). While a certain portion of the energy of UV quanta is expended for breaking bonds of electrons with atoms and of atoms with molecules, the remaining energy is taken by the products of reactions (mainly by electrons). This kinetic energy of photoprocess products finally results in the gas heating. Modeling these processes in detail is an individual and complex problem. Here, we model no processes of ionization, dissociation, and recombination, nor other processes of the microphysics responsible for heating and cooling atmospheres. Instead of that, we use a modeling heating function that simulates heating by
the gas photoionization 
\begin{equation}
\Gamma=3.57\, \kappa F_0\, \exp\left\{-\tau-\dfrac{1}{\tau}\right\},
\label{modelheat}
\end{equation}
where $F_0$ --- the radiation flux, $\kappa$ --- absorption coefficient, $\tau = \kappa \Sigma$ --- optical thickness, $\Sigma$ --- integral ray density calculated from the outer atmospheric boundary to the given point. Within this modeling description, we assume that the gas is mainly heated near $\tau=1$. Indeed, for
$\tau \ll 1$ the gas is completely ionized and does not participate in heating; for $\tau \gg 1$, the radiation does not penetrate and does not heat the gas. Note
that our optical thickness $\tau$ depends on the ray density of both neutral and ionized components, instead of that depending on the ray density of the neutral gas
only. As far as the atmosphere flows out, the gas initially located in the heating region moves outward, expands, and becomes transparent to the UV radiation. The factor $3.57$ normalizes the integral over the total optical thickness. The considered approach is a phenomenological approximation; for a more substantiated description, one must calculate the content of neutral and ionized gases within a more complex (aeronomic) model. In our model, the gas can be cooled only at the expense of making work, i.e., owing to expansion.

The model described above does not require any detailed modeling of kinetic processes, but provides qualitatively correct patterns of outflow (as we further see) and is rather easy. The latter is important for testing the calculations on the atmosphere dynamics. The parameters of the introduced heating function are $F_0$ and $\kappa$. The parameter $F_0$ describes the total flux of the UV energy absorbed by the atmosphere. The parameter $\kappa$ determines the spatial location of the region of efficient heating. Taking into account our above conclusion on the importance of external boundary conditions for modeling the atmosphere outflow, we
must indicate an additional key parameter for this problem, namely, the pressure on the moving outer boundary $P_{\rm out}$.

\begin{table}

\caption{Atmospheric parameters of the basic model equipped with the phenomenological heating function. }
\begin{tabular}{l|l|l}
\hline
Parameter		& Notation	& Value \\
\hline
Planet mass			& $M$			& 0.07~$M_J$ \\
Planet radius			& $a$			& 0.35~$R_J$ \\
Initial temperature		& $T_0$			& 750~K \\
UV radiation flux	 	& $F_0$			& $10^3$~erg s$^{-1}$ cm$^{-2}$ \\
Absorption coefficient		& $\kappa$		& $1.5\times 10^6$ cm$^2$ g$^{-1}$ \\
Density at the inner boundary	& $\rho_0$		& $10^{-10}$ g cm$^{-3}$ \\
Pressure at the outer boundary	& $P_{\rm out}$	& $10^{-6}$ dyne cm$^{-2}$ \\
\hline
\end{tabular}
\label{neptune_par}
\end{table}

Let us consider the atmosphere of the planet whose parameters are close to those of the warm Neptune GJ 436b (see Table~\ref{neptune_par}). We take the mass and radius of the planet (from the site http://www.exoplanet.eu) with the initial temperature and the flux being the same as those in~\cite{2018MNRAS.481.5315S}. We take the absorption coefficient as the ratio of the ionization cross-section for hydrogen atoms $\sigma_{\rm H}$ at the wavelength of 912~\AA\, to the hydrogen atom mass
$m_{\rm H}$, namely, $\kappa=\sigma_{\rm H}/m_{\rm H}$. The basic model implies that the atmosphere consists of the molecular hydrogen, with its adiabatic index being 7/5. For the initial time, the atmosphere is at the hydrostatic equilibrium, with the pressure at the outer boundary being equal to the external pressure $P_{\rm out}$.

 \begin{figure*}  
 \centering
 \includegraphics[width=2\columnwidth]{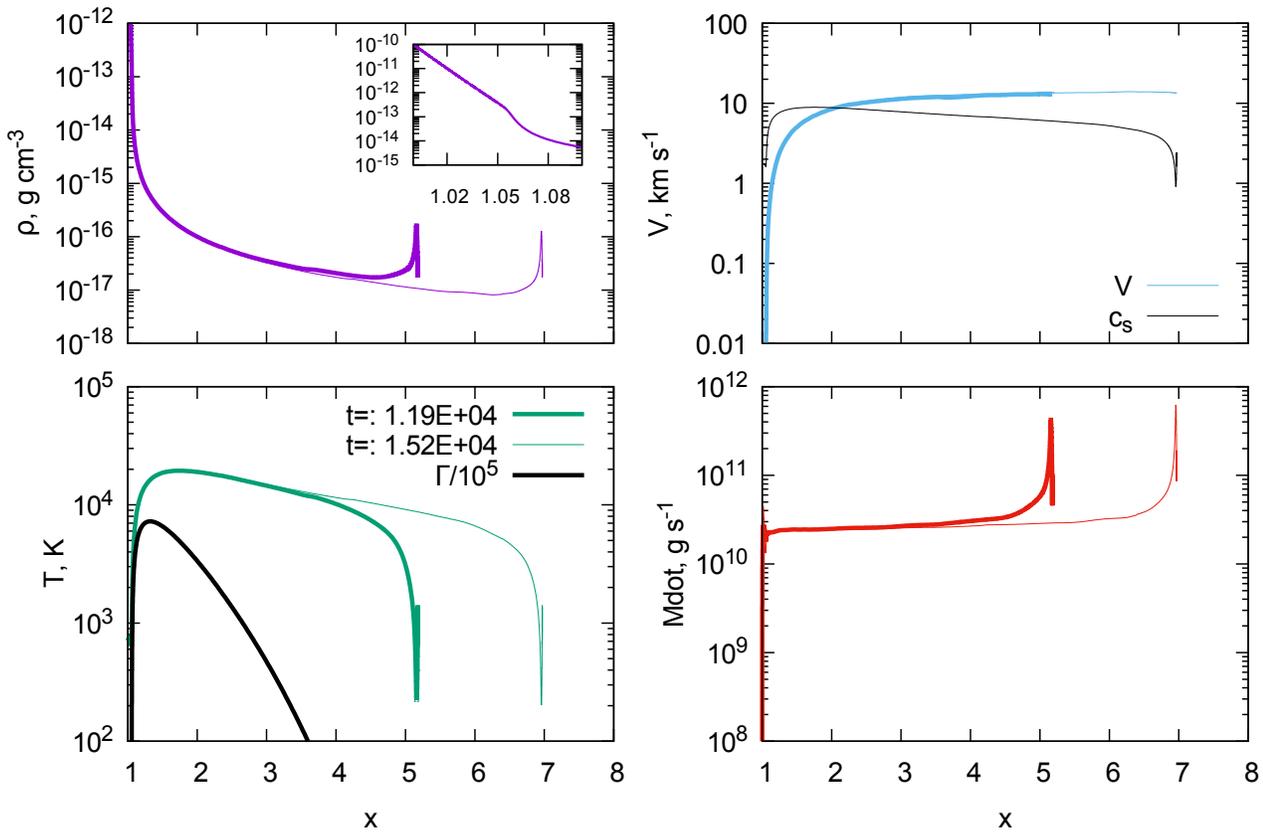}
 \caption{Basic atmospheric outflow model equipped with the simple heating function.}
 \label{neptune-base}
 \end{figure*}

Fig.~\ref{neptune-base} presents distributions of the density, velocity, temperature, and outflow rate for the basic model at two times. For the time of $1.52\times10^4$~s, the outer atmospheric boundary reaches $\sim 7$~planet radii, while the outflow rate within $x<6.5$ is close to constant and equals
$3\times10^{10}$~g/s. In addition, for both times presented in Fig.~\ref{neptune-base} ($1.19 \times 10^4~\text{s}$ and $1.52 \times 10^4~\text{s}$) the mass transfer rates in inner regions of the outflow are identical, and this indicates a steady regime of the outflow. The atmosphere can be tentatively divided
into several regions. In the innermost region ($x<1.06$), the atmosphere is close to hydrostatic equilibrium, where the density rapidly decreases with
distance and the velocity approaches zero. Within the region $1.06<x<2$, the atmosphere is intensively heated, and this is indicated with the distribution $\Gamma(x)$ shown by the black curve in the left lower panel of Fig.~\ref{neptune-base}. In the region of atmospheric heating, the out-flow velocity rapidly increases with distance. The density distribution changes in this region, and the profile becomes flatter (see the inset in the left upper panel of Fig.~\ref{neptune-base}. The temperature reaches its maximum of $2\times 10^{4}$~K inside the heating region. Within the region $2<x<7$, where the atmospheric heating is weak, the outflow velocity gradually increases with distance. Here, the temperature gradually decreases with distance due to the adiabatic expansion. In outermost regions of the outflow, there are significant gradients of all physical quantities. These features result from the initial and boundary conditions of our modeling. Indeed, the atmosphere at the initial time is in hydrostatic equilibrium with a constant temperature and monotonic density distribution. However, at later times, the atmosphere is heated by the external radiation, which raises the temperature and enhances the pressure in the zone of radiation absorption (located at a certain distance from the boundary). This results in the expansion of the envelope and compression of the near-boundary layers of lower temperatures. This, in turn, results in density peaks located near the boundary. These features do not affect the distributions of physical quantities in inner layers of the atmosphere. Distributions of the gas velocity and the speed of sound are shown in the right upper panel of Fig.~\ref{neptune-base}. We can see that the flow becomes supersonic at the dis-
tance $x=2$. Note that for the outflow rate obtained, the planet mass will not considerably change over the cosmological time.

\begin{figure*}
\centering
\includegraphics[width=0.66\columnwidth]{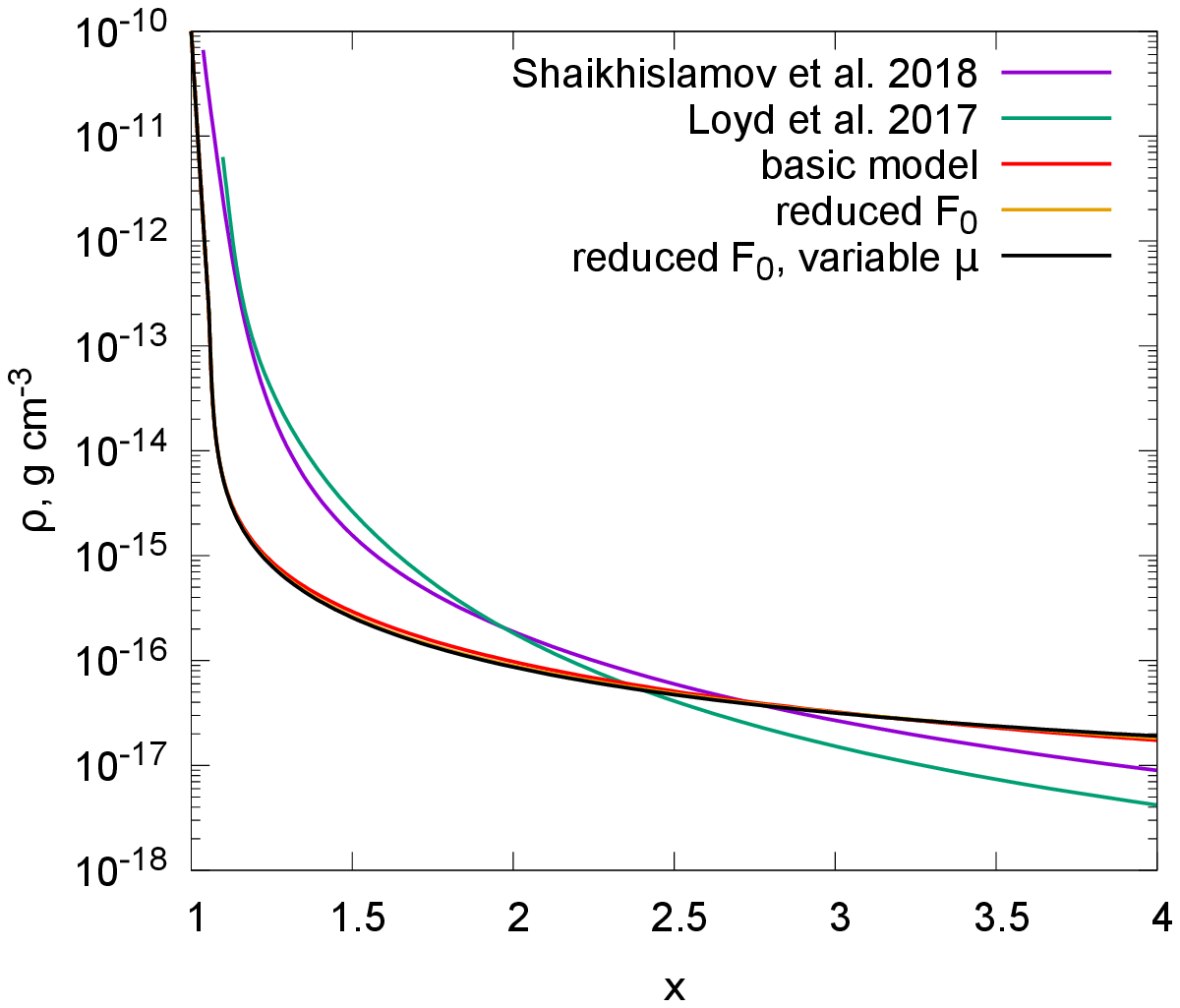}
\includegraphics[width=0.66\columnwidth]{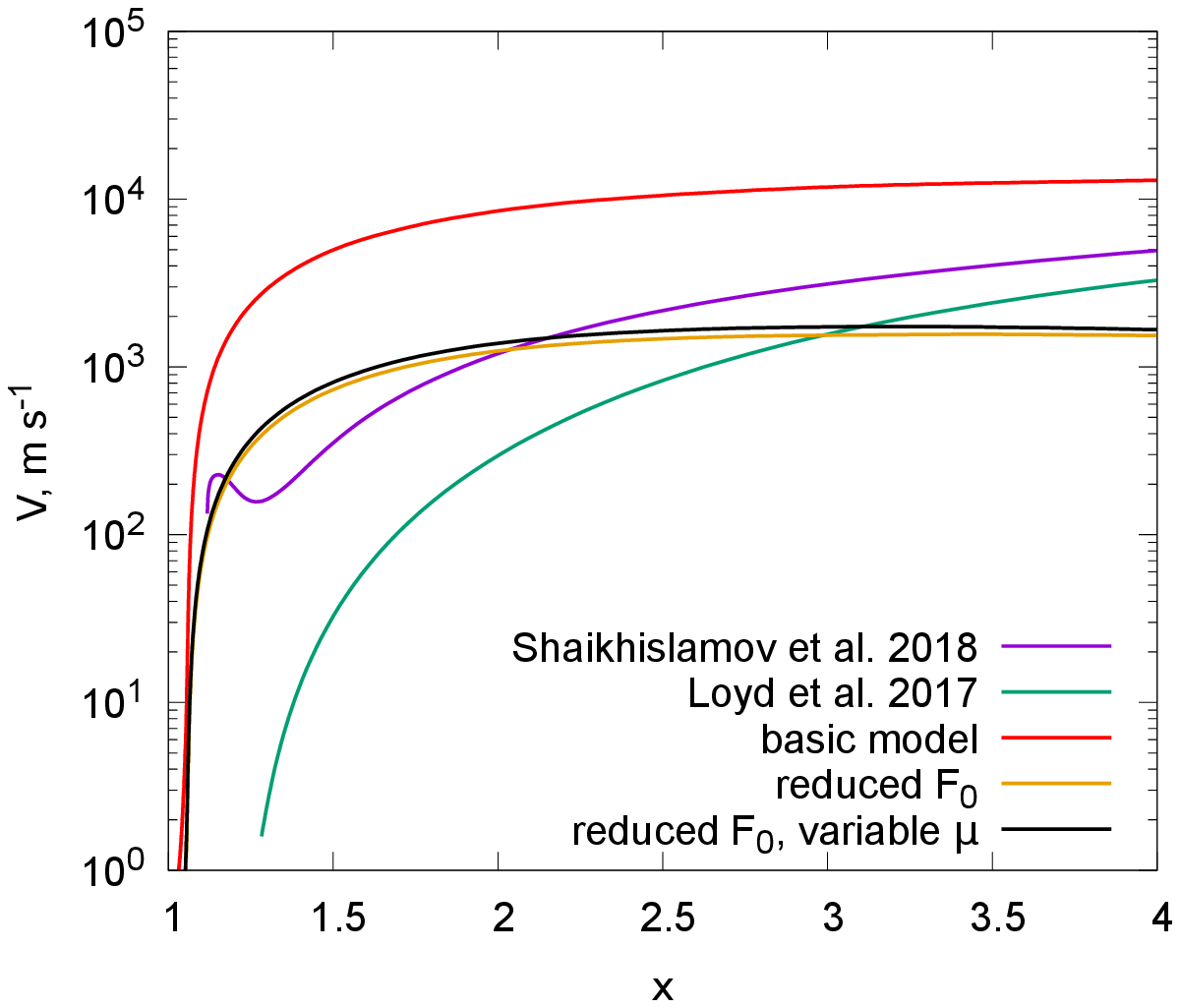}
\includegraphics[width=0.66\columnwidth]{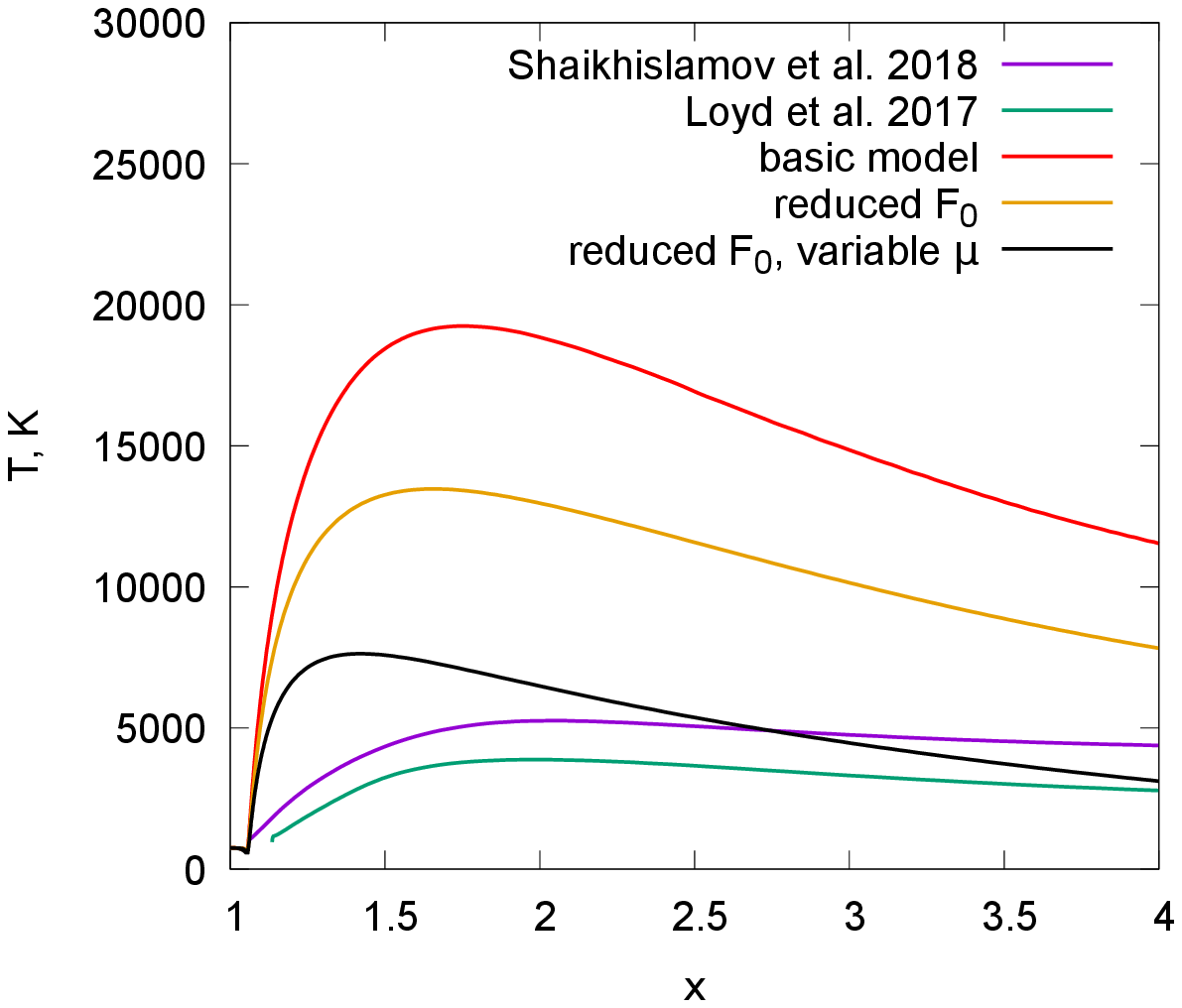}
\caption{Basic outflow model compared with other aeronomic models. Atmospheric parameters for the basic model with the simple heating function are shown red, for the model with radiation flux $F_0$, reduced by an order of magnitude yellow, for the model with the reduced UV flux and varying molecular weight black, 
for the aeronomic model~\citep{2018MNRAS.481.5315S} violet, and  green for the aeronomic model~\citep{2017ApJ...834L..17L}.}
\label{earonomic}
\end{figure*}

Fig.~\ref{earonomic} compares the obtained distributions with aeronomic models used for GJ~436b. The calculations presented in~\cite{2018MNRAS.481.5315S} and \cite{2017ApJ...834L..17L} are shown violet and green, respectively. Along with hydrodynamic processes, the cited calculations consider the ionization, dissocia-
tion, recombination, and other photochemical processes occurring with H$_2$, H, He, H$^+$, H$^+_2$, H$^+_3$, etc. In outline, our approach reproduces the morphology of density, velocity, and temperature distributions obtained in aeronomic models, but with considerable quantitative differences. In particular, the gas-
dynamic model with the phenomenological heating function displays the temperature overestimated by several times (see red curves indicated as ``basic model''). The efficiency of the atmospheric heating by UV quanta was shown to be about $0.1$~\cite{2014A&A...571A..94S}. We take this into account by an additional calculation using the heating rate $F_0$ of one tenth of that used for basic calculations.

\begin{figure*}
\centering
\includegraphics[width=0.95\textwidth]{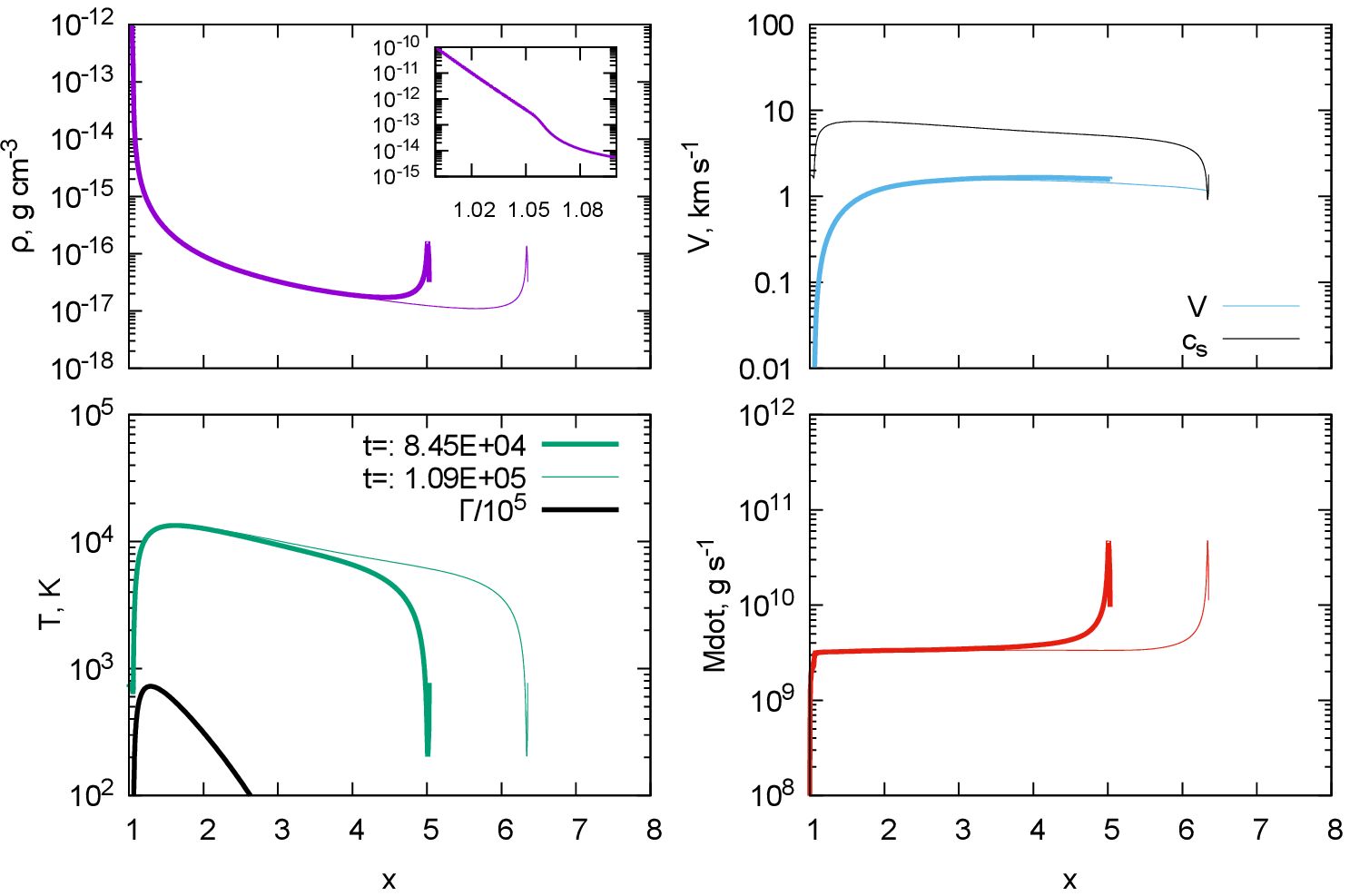}
\caption{Atmospheric outflow model with the radiation flux $F_0$ reduced by an order of magnitude.}
\label{neptune-F0}
\end{figure*}

Fig.~\ref{neptune-F0} presents calculations executed for the model with the reduced heating rate. Morphological features of distributions obtained here are close to
those calculated in the basic model, but the velocity and the outflow rate become by an order of magnitude lower than those of the basic model. The maximum temperature is reduced to about 13\,000~K, (see yellow curves shown in~\ref{earonomic} and indicated as ``reduced $F_0$'') The flow becomes subsonic everywhere. Though the temperature is reduced, this remains by 2-3 times higher than that obtained in aeronomic models.

In order to correctly reproduce the physics of the atmosphere (including temperature distributions), one must consider a great number of physical processes. In particular, aeronomic models must neces-sarily calculate the degree of dissociation and ionization of basic molecules. This provides correct calculations for the field of UV radiation penetrating into lower layers of the atmosphere. In addition, the dissociation and ionization elevate the number density of free particles. A local increase in particle number densities (at a constant temperature) results in an increase in pressure according to equation~\eqref{eq-state}. Thus, the pressure in the zone of absorption of UV quanta increasesdue to both the increase in temperature and the dissociation/ionization of the molecular hydrogen. Let us demonstrate an influence of the latter effect on the thermal structure of the atmosphere in frameworks of the gas-dynamic model used. For this purpose, we assume that the average molecular mass (the ratio of the mean mass of particles to the proton mass) depends on the optical thickness $\tau$ as follows:
\begin{equation}
\mu=\dfrac{1}{2}+\dfrac{3\tau}{2(1+\tau)}.
\label{varmu}
\end{equation}
This expression describes a gradual transition of $\mu$ from $1/2$ of the optically thin medium (where hydrogen is completely dissociated and ionized) to the mass
of $2$ for the highest optical thickness (where hydrogen is completely molecular). Thus, we vary the number of degrees of freedom $i$ and the adiabatic index $\gamma$ using the expressions $i=3+\dfrac{4}{3}(\mu-\dfrac{1}{2})$, $\gamma=\dfrac{i+2}{i}$. Note that using spatially varying $\mu$ and $\gamma$ in gas-dynamic codes, we conserve the specific thermal energy under transformations between the pressure, temperature, and thermal energy.

\begin{figure*}
 \centering
 \includegraphics[width=0.95\textwidth]{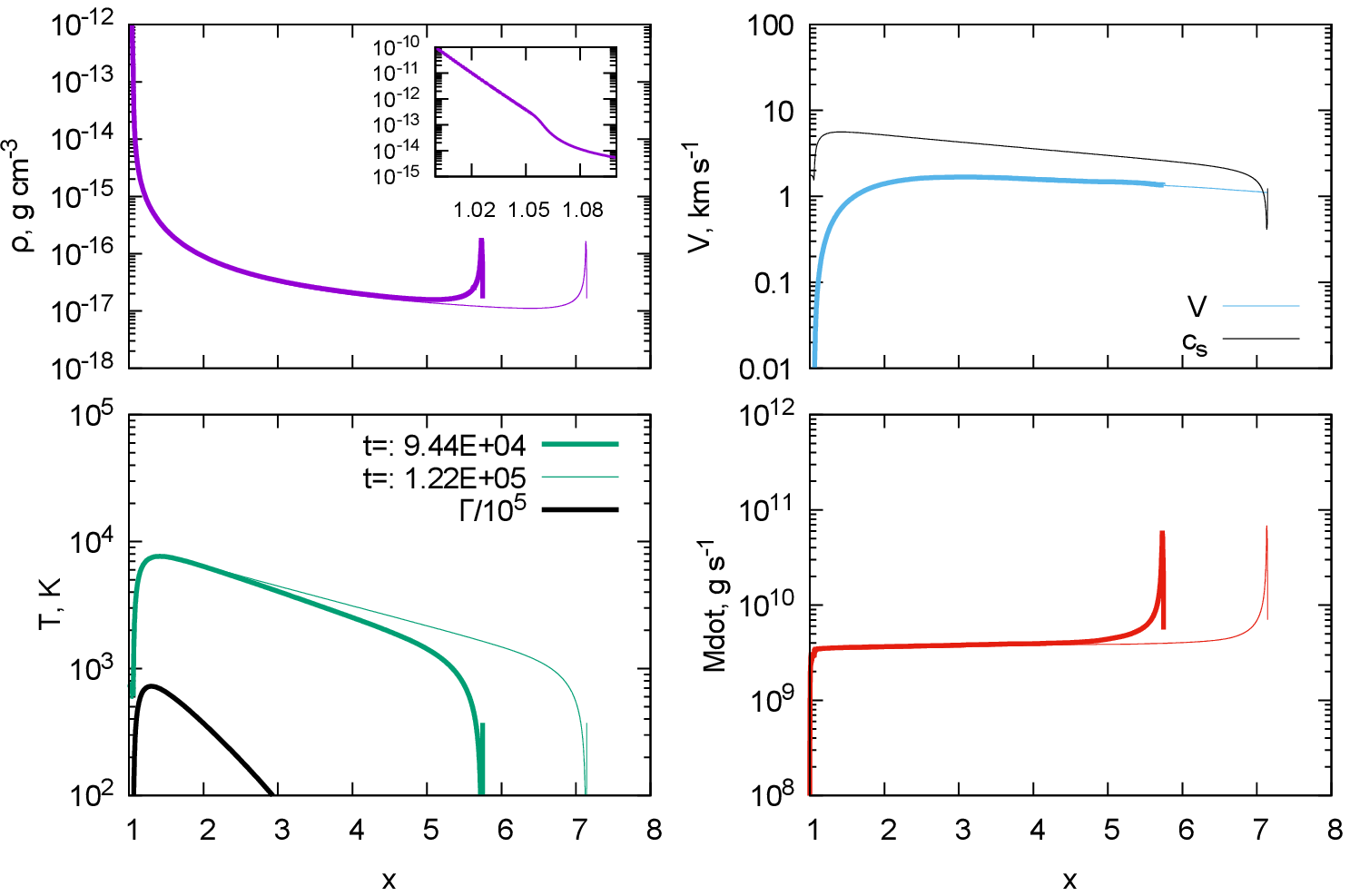}
 \caption{Atmospheric outflow model with the reduced UV flux and varying molecular weight.}
 \label{neptune-ion}
 \end{figure*}

The calculations for the model with the mean molecular weight and adiabatic index depending on the optical thickness along with the reduced flux of radiation are indicated as ``reduced $F_0$, variable $\mu$'' and shown black in Figs.~\ref{earonomic} and ~\ref{neptune-ion}.

A key feature of this model is that the maximum temperature reduces to $\approx 7\,000$~K, which is comparable to that of aeronomic models. Note also that the rate of mass losses ($3\times10^9$~g/s) does not change compared with that for the model of constant $\mu$, $\gamma$ with reduced $F_0$.

It is helpful to compare the obtained rate of outflow $\dot{M}$ with the estimate proposed in~\cite{1981Icar...48..150W}:
\begin{equation}
F = \fr{S r_1^2 r_0}{G M m},
\label{F-Watson}
\end{equation}
where we save the original notations, namely, $F$~--- the flux of escaping particles per unit time in unit spatial angle,  $S$~--- total heating rate (corresponds to $F_0$ in our notations)), $r_1$~--- level of unit optical thickness, $r_0$~--- planet radius, $m$~--- mass of escaping particles (hydrogen atom mass in our case), 
$M$~--- planet mass. Taking $r_1 \approx r_0 = a$, $F_0 = 10^2$~erg s$^{-1}$ cm$^{-2}$ and multiplying equation~\eqref{F-Watson} by $4 \pi$ and by hydrogen atom mass, we find an estimate for the atmosphere outflow rate $\dot{M} \approx 2 \times 10^{9}~\text{g/s}$. This rate well agrees with that of our modeling.

It is quite evident that the atmosphere dynamics associated with heating by external radiation is a much more complex process than isothermal or adiabatic
winds. This puts forward numerical methods solving the problems on dynamics of planetary atmospheres. The models presented in this section are mainly illustrative and methodical. Nevertheless, the proposed phenomenological description of heating can be used for approximate models of the atmosphere outflow neglecting the ionization-chemical processes. The proposed models can also be used for testing the gas-dynamic methods of aeronomic modeling.

\section{Conclusions}

Our work is rather introduction into a rapidly growing branch of modern astrophysics dealing with models on the outflow of planetary atmospheres. It is quite difficult to find analytical solutions even in simple approximations, so numerical methods of modeling the atmosphere dynamics come to the foreground. The gas-dynamic method, which must satisfy rather high requirements arising from great gradients of physical quantities and supersonic-flow regimes in combination with modeling quasi-equilibrium states, is among the key elements of such models. Here, we demonstrate examples of testing the gas-dynamic module, which is widely used in the software assigned for modeling the physical structure of exoplanet atmospheres~\cite{2017ARep...61..387I,2018MNRAS.tmp..609I}. It is shown that with proper boundary conditions the gas-dynamic method reproduces the analytical solutions found for isothermal atmospheres.

Modern models of the atmosphere outflow (so-called aeronomic models) include calculations of the ionization, chemical, and thermal structure of atmospheres. Within these models, we can study the finest physical effects and approach detailed interpretations of observational data (see, for example,~\cite{2018MNRAS.481.5315S,2020A&A...639A.109S}). However, such models are very complex, and these can hardly be used for educational purposes. Our work presents an example showing that the general outflow of the atmosphere exposed to stellar UV heating can be described by elementary gas-dynamic models equipped with phenomenological heating functions.
Within this model, we have presented the effects exerted by heating rates and varying molecular weights on the flow characteristics. This model can be used for
methodical purposes and for testing gas-dynamic packages of aeronomic models, as well as for approximate calculations carried out within models neglecting the detailed chemical and ionization structure of atmospheres.

\section*{Funding}
P.B.I. (Section 2) was supported by the Russian Science Foundation (project 18-12-00447). V.I.Sh. and E.S.K. (Sections 1 and 3) were supported by a grant from the Government of the Russian Federation for Scientific Research under the leadership of leading scientists within the Project ``Investigation of Stars with Exoplanets'' (contract no. 075-15-2019-1875).

\bibliographystyle{mn2e} 
\bibliography{envelope}

\end{document}